\documentclass[12pt]{article}

\usepackage[totalwidth=460pt,totalheight=615pt]{geometry}
\usepackage{amsfonts,latexsym,amssymb,amsmath,graphicx,accents,slashed,subfigure}
\usepackage[T1]{fontenc}
\usepackage[hidelinks]{hyperref}

\global\arraycolsep=1truept

\numberwithin{equation}{section}

\begin{document}

\bigskip \phantom{C}

\vskip1truecm

\begin{center}
{\huge \textbf{Fakeons, Microcausality}}

\vskip.4truecm

{\huge \textbf{And The Classical Limit}}

\vskip.6truecm

{\huge \textbf{Of Quantum Gravity}}

\vskip1truecm

\textsl{Damiano Anselmi}

\vskip .1truecm

\textit{Dipartimento di Fisica ``Enrico Fermi'', Universit\`{a} di Pisa}

\textit{and INFN, Sezione di Pisa,}

\textit{Largo B. Pontecorvo 3, 56127 Pisa, Italy}

damiano.anselmi@unipi.it

\vskip2truecm

\textbf{Abstract}
\end{center}

We elaborate on the idea of fake particle and study its physical
consequences. When a theory contains fakeons, the true classical limit is
determined by the quantization and a subsequent process of \textquotedblleft
classicization\textquotedblright . One of the major predictions due to the
fake particles is the violation of microcausality, which survives the
classical limit. This fact gives hope to detect the violation
experimentally. A fakeon of spin 2, together with a scalar field, is able to
make quantum gravity renormalizable while preserving unitarity. We claim
that the theory of quantum gravity emerging from this construction is the
right one. By means of the classicization, we work out the corrections to
the field equations of general relativity. We show that the finalized
equations have, in simple terms, the form $\langle F\rangle =ma$, where $%
\langle F\rangle $ is an average that includes a little bit of
\textquotedblleft future\textquotedblright .

\vfill\eject

\section{Introduction}

\setcounter{equation}{0}

A theory of quantum gravity has been recently formulated \cite{LWgrav}, by
means of a new quantization prescription that turns certain poles of the
free propagators into fake particles, or \textquotedblleft
fakeons\textquotedblright , which are not asymptotic states and must be
projected away from the physical spectrum. The action of the theory is
interpreted in a radically new way and some of its physical predictions are
quite unexpected, even at the classical level. In particular, the physical
space of configurations is a proper subspace of the whole space of
configurations. The restriction follows from the projection on the space of
states that is required to make sense of the theory at the quantum level. In
this paper, we study how these features of quantum gravity affect the
classical limit. In particular, we work out the corrections to the field
equations of general relativity and investigate their major prediction,
which is the violation of microcausality.

The \textit{interim} classical action is 
\begin{equation}
S_{\text{QG}}(g,\Phi )=-\frac{1}{2\kappa ^{2}}\int \sqrt{-g}\left[ 2\Lambda
_{C}+\zeta R+\alpha \left( R_{\mu \nu }R^{\mu \nu }-\frac{1}{3}R^{2}\right) -%
\frac{\xi }{6}R^{2}\right] +S_{\mathfrak{m}}(g,\Phi ),  \label{SQG}
\end{equation}%
where $\alpha $, $\xi $, $\zeta $ and $\kappa $ are real positive constants,
while $\Lambda _{C}$ can be positive or negative, $\Phi $ are the matter
fields and $S_{\mathfrak{m}}$ is the covariantized action of the standard
model, or one of its extensions, equipped with the nonminimal couplings that
are compatible with renormalizability. The reduced Planck mass is $\bar{M}_{%
\text{Pl}}=M_{\text{Pl}}/\sqrt{8\pi }=\sqrt{\zeta }/\kappa $. Here and
below, the integration measure $\mathrm{d}^{4}x$ is understood.

In addition to the matter fields, the theory describes the graviton, a
scalar $\phi $ of squared mass $m_{\phi }^{2}=\zeta /\xi $\ and a spin-2
fakeon $\chi _{\mu \nu }$ of squared mass $m_{\chi }^{2}=\zeta /\alpha $
(neglecting a small correction due to the cosmological constant). These
fields can be introduced by means of auxiliary fields and simple field
redefinitions. We obtain an equivalent form of the interim classical action,
which reads \cite{absograv} 
\begin{equation}
\mathcal{S}_{\text{QG}}(g,\phi ,\chi ,\Phi )=S_{\text{H}}(g)+S_{\chi
}(g,\chi )+S_{\phi }(\tilde{g},\phi )+S_{\mathfrak{m}}(\tilde{g}\mathrm{e}%
^{\kappa \phi },\Phi ),  \label{SQG2}
\end{equation}%
where~$\tilde{g}_{\mu \nu }=g_{\mu \nu }+2\chi _{\mu \nu }$ and 
\begin{eqnarray}
&&S_{\text{H}}(g)=-\frac{\zeta }{2\kappa ^{2}}\int \sqrt{-g}R,\qquad S_{\phi
}(g,\phi )=\frac{3\zeta }{4}\int \sqrt{-g}\left[ \nabla _{\mu }\phi \nabla
^{\mu }\phi -\frac{m_{\phi }^{2}}{\kappa ^{2}}\left( 1-\mathrm{e}^{\kappa
\phi }\right) ^{2}\right] ,  \notag \\
&&S_{\chi }(g,\chi )=S_{\text{H}}(\tilde{g})-S_{\text{H}}(g)-2\int \chi
_{\mu \nu }\frac{\delta S_{\text{H}}(\tilde{g})}{\delta g_{\mu \nu }}+\frac{%
\zeta ^{2}}{2\alpha \kappa ^{2}}\int \left. \sqrt{-g}(\chi _{\mu \nu }\chi
^{\mu \nu }-\chi ^{2})\right\vert _{g\rightarrow \tilde{g}}.  \label{ss}
\end{eqnarray}%
We have written these expressions for $\Lambda _{C}=0$, which is sufficient
for most purposes of this paper. The formulas for $\Lambda _{C}\neq 0$ can
be found in ref. \cite{absograv}.

The first thing to point out is that (\ref{SQG}) and (\ref{SQG2}) do not
encode the true classical theory, which is why we have called them
\textquotedblleft interim\textquotedblright\ classical actions. Indeed, they
are still unprojected. However, the projection comes from the quantization,
so we must first quantize the theory and then \textit{classicize} it back,
which means investigate the classical backlash of the quantization. Only at
the end of this procedure, we obtain the action that describes the classical
limit, which we call \textit{finalized classical action}. The classicization
has to be understood perturbatively, since it is inherited from quantum
field theory, which is formulated perturbatively. The interim classical
action and the finalized classical action coincide in the absence of gravity
(for example, in the case of the standard model in flat space, where the
projection is trivial), but do not coincide when quantum gravity is present.

The second major observation is that the fakeons induce violations of
microcausality, which survive the classical limit. This fact opens the
possibility to investigate such violations, and other predictions of the
theory, beyond the scattering processes and possibly beyond the perturbative
expansion, by studying the solutions of the projected classical field
equations.

We recall that if the action (\ref{SQG}) is quantized following the standard
procedures (which means using the Feynman prescription for all the fields),
the Stelle theory is obtained \cite{stelle}, where $\chi _{\mu \nu }$ is a
ghost. This option is not acceptable, because there exists no consistent
projection to get rid of a ghost.

Several approaches to the problem of quantum gravity have been proposed in
the past decades, such as string theory \cite{string}, loop quantum gravity 
\cite{loop}, holography and the AdS/CFT\ correspondence \cite{ads}, lattice
gravity \cite{latticeg}, asymptotic safety \cite{asafety}. It is beyond the
scope of this paper to give a comprehensive account of the vast literature
on the subject. Nevertheless, it is useful to make a brief comparison
between the properties of our solution and those of the most popular
alternatives, to highlight some key features and emphasize the reasons why
we believe that our proposal is the right one.

The approach we adopt is very conservative and follows the guidelines of
high-energy particle physics. Its basic principles (unitarity, locality and
renormalizability) are the same that worked successfully for the standard
model.

Like the standard model, our theory of quantum gravity is a quantum field
theory and admits a perturbative expansion in terms of Feynman diagrams. It
can be straightforwardly coupled to the standard model, as shown in formulas
(\ref{SQG}) and (\ref{SQG2}). The calculations are doable and demand an
effort that is comparable to the one required by familiar computations in
particle physics (see refs. \cite{absograv,UVQG}).

The other approaches we have mentioned have different motivations and goals.
Each of them has interesting features, but also important drawbacks. For
example, string theory faces criticisms because of its lack of predictivity,
originated by the landscape of 10$^{500}$ or so false vacua \cite{landscape}%
. Moreover, perturbative calculations in string theory are much more
difficult than those of quantum field theory (apart from special cases),
since integrals on Riemann surfaces require mathematics that is not
completely understood. To some extent, loop quantum gravity is even more
challenging, from the mathematical point of view. The AdS/CFT correspondence
has a quantum field theorical side, which is however strongly coupled and
requires the application of nonperturbative methods. Lattice gravity, like
lattice gauge theory, is a numerical approach, useful for a variety of
purposes, but with limited power to shed light on conceptual aspects. The
asymptotic-safety program is fully field theoretical, like our approach.
Nevertheless, it is not perturbative, since it requires the existence of an
interacting ultraviolet fixed point.

We think that our solution tops the competitors in calculatibity,
predictivity and falsifiability. This also means that it is rather rigid.
Indeed, it contains only two new parameters (the masses of $\phi $ and $\chi
_{\mu \nu }$), so there is not much room for adjustments in the case of
discrepancies with the data. 

Dealing with quantum gravity, it is generically assumed that we must be
prepared to accept some profoundly new understanding of spacetime at the
microscopic level. In our case, the conceptual upset emerges from the theory
itself and is precisely the violation of microcausality, i.e. the fact that
space and time, past, present and future, cause and effect lose physical
meaning at energies larger than the lightest fakeon mass. Our present
knowledge does leave room for this prediction to be accurate. Actually, from
the theoretical and experimental points of view, we lack compelling reasons
to believe that causality, no matter how it is defined, should hold down to
arbitrarily small distances.

It should be recalled that causality is not well understood in quantum field
theory, to the extent that a formulation that corresponds to the intuitive
notion is missing \cite{diagrammar}. The best we have are off-shell
formulations, such as Bogoliubov's diagrammatic condition \cite{bogoliubov},
which implies, among other things, that fields commute at spacelike
separated points. However, Bogoliubov's condition cannot be formulated as a
constraint on the $S$ matrix, because it is not possible to accurately
localize spacetime points working with relativistic wave packets that
correspond to on-shell particles. These observations explain why
microcausality has not been treated so far as a fundamental requirement,
maybe in anticipation that it was going to be renounced eventually. 

The paper is organized as follows. In section \ref{quanti} we recall what a
fakeon is and elaborate on the concept. In section \ref{predi} we study the
main physical implications due to a fakeon at the level of scattering
processes, such as the violation of microcausality. In section \ref{toy}\ we
study how the violations of microcausality survive the classical limit in a
toy higher-derivative model. In section \ref{class} we classicize quantum
gravity. In particular, we work out the finalized classical action and study
its field equations.\ Section \ref{conclu} contains the conclusions.

\section{Quantization}

\setcounter{equation}{0}\label{quanti}

In this section, we elaborate on the idea of fake particle, or fakeon,
introduced in ref. \cite{LWgrav}. Unitarity is crucial for the discussion,
so we begin by recalling the optical theorem%
\begin{equation}
2\hspace{0.01in}\mathrm{Im}T=T^{\dag }T,  \label{optical}
\end{equation}%
which follows from the unitarity equation $S^{\dag }S=1$, once the $S$
matrix is written as $1+iT$. Let $V$ denote the space of physical states.
The transition amplitude $\mathcal{M}(a|b)$ between an initial state $%
|a\rangle \in V$ of total momentum $P_{a}$ and a final state $|b\rangle \in V
$ of total momentum $P_{b}$ is related to the $T$ matrix element by the
identity%
\begin{equation*}
\langle b|T|a\rangle =(2\pi )^{4}\delta ^{(4)}(P_{a}-P_{b})\mathcal{M}(a|b).
\end{equation*}%
Moreover, $i\mathcal{M}(a|b)$ is the sum of the connected, amputated
diagrams whose external legs are determined by $|a\rangle $ and $|b\rangle $%
. Taking $|b\rangle =|a\rangle $ and inserting a complete set of orthonormal
states $|n\rangle \in V$, equation (\ref{optical}) implies 
\begin{equation}
2\hspace{0.01in}\mathrm{Im}\langle a|T|a\rangle =\sum_{|n\rangle \in
V}|\langle n|T|a\rangle |^{2},  \label{o2}
\end{equation}%
i.e. the total cross section for production of all final states is
proportional to the imaginary part of the forward scattering amplitude. A
version of the optical theorem that holds diagram by diagram is provided by
the so-called cutting equations \cite{cuttingeq}. A cutting equation
expresses the real part of a diagram (which is equal to minus the imaginary
part of its contribution to the amplitude $\mathcal{M}$) as a sum of
\textquotedblleft cut diagrams\textquotedblright , where the contributions
of $\langle n|T|a\rangle $ and $\langle a|T^{\dagger }|n\rangle $ stand to
the left and right sides of the cuts, respectively. The simplest cutting
equations are 
\begin{eqnarray}
2\hspace{0.01in}\text{Im}\left[ (-i)%
\raisebox{-1mm}{\scalebox{2}{$\rangle
\hspace{-0.075in}-\hspace{-0.07in}\langle$}}\,\right] =%
\raisebox{-1mm}{\scalebox{2}{$\rangle
\hspace{-0.075in}-\hspace{-0.14in}\slash\hspace{-0.015in}\langle$}} &=&\int 
\mathrm{d}\Pi _{f}\hspace{0.01in}\left\vert \raisebox{-1mm}{\scalebox{2}{$%
\rangle\hspace{-0.035in}-$}}\right\vert ^{2},  \label{cutd} \\
2\hspace{0.01in}\text{Im}\left[ (-i)\raisebox{-1mm}{\scalebox{2}{$-%
\hspace{-0.065in}\bigcirc\hspace{-0.065in}-$}}\right] =\hspace{0.01in}%
\raisebox{-1mm}{\scalebox{2}{$-\hspace{-0.065in}\bigcirc\hspace{-0.16in}%
\slash\hspace{0.015in}-$}} &=&\int \mathrm{d}\Pi _{f}\hspace{0.01in}%
\left\vert \raisebox{-1mm}{\scalebox{2}{$-\hspace{-0.035in}\langle$}}%
\right\vert ^{2},  \label{cutdd}
\end{eqnarray}%
where the integrals are over the phase spaces $\Pi _{f}$ of the final states 
\cite{peskin}.

Now, consider the propagator%
\begin{equation}
G(p,m)=\frac{1}{p^{2}-m^{2}}.  \label{prop}
\end{equation}%
If we endow it with the Feynman prescription ($p^{2}\rightarrow
p^{2}+i\epsilon $), we obtain%
\begin{equation}
G_{+}(p,m,\epsilon )=\frac{1}{p^{2}-m^{2}+i\epsilon },  \label{propF}
\end{equation}%
which describes a particle of mass $m$. The identity (\ref{cutd}) implies $%
\mathrm{Im}[-P]\geqslant 0$, if the vertices are assumed to be equal to $-i$
and $P$ is the propagator of the intermediate line on the left-hand side.
Specifically, $P=G_{+}$ gives 
\begin{equation}
\mathrm{Im}\left[ -\frac{1}{p^{2}-m^{2}+i\epsilon }\right] =\pi \delta
(p^{2}-m^{2}).  \label{opti}
\end{equation}

What happens if we multiply (\ref{propF}) by a minus sign? If we do not do
anything else, we obtain a ghost, since $P=-G_{+}$ satisfies $\mathrm{Im}%
[-P]\leqslant 0$, which violates the optical theorem. However, if we also
replace $+i\epsilon $ with $-i\epsilon $, the right-hand side of (\ref{opti}%
) does not change, 
\begin{equation}
\mathrm{Im}\left[ \frac{1}{p^{2}-m^{2}-i\epsilon }\right] =\pi \delta
(p^{2}-m^{2}),  \label{opti2}
\end{equation}%
and the optical theorem remains valid. The moral of the story is that we can
in principle have both propagators%
\begin{equation*}
G_{\pm }(p,m,\epsilon )=\pm \frac{1}{p^{2}-m^{2}\pm i\epsilon },
\end{equation*}%
since both fulfill the identity (\ref{cutd}).

Nevertheless, this is not what normally occurs in quantum field theory, at
least within the same loop integral. The reason is that the presence of both 
$G_{+}$ and $G_{-}$ originates bad nonlocal divergences at $\epsilon \neq 0$ 
\cite{ugo} and even worse problems for $\epsilon \rightarrow 0$.

A place where both propagators do appear in the same diagram are the cutting
equations already mentioned: one side of a cut diagram is built with the
propagators $G_{+}$ and the other side is built with the propagators $G_{-}$%
. For the reasons just recalled, a cut diagram gives a loop integral that
can be badly divergent, because it contains both $G_{+}$ and $G_{-}$.
However, the sum of the cut diagrams is well defined, because, by the
optical theorem, it is equal to the real part of an uncut diagram, which in
turn is the sum of a diagram built only with $G_{+}$ plus a diagram built
only with $G_{-}$.

Is there any hope to have $G_{\pm }$ coexist consistently in the same loop
integral, i.e. an ordinary, uncut diagram? The first thing to note is that
we should not integrate directly on Minkowski spacetime, to avoid the
nonlocal divergences of ref. \cite{ugo}. The only alternative is to come
from Euclidean space by means of the Wick rotation \cite{LWformulation}. It
turns out that the Wick rotation is not analytic. However, this difficulty
is not serious enough to prevent us from moving further. 
\begin{figure}[t]
\begin{center}
\includegraphics[width=10truecm]{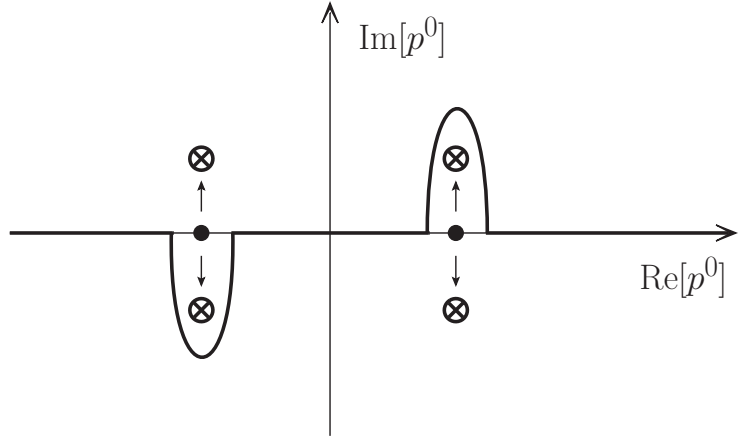}
\end{center}
\caption{Splitting the poles located on the real axis. The dots denote the
poles of (\protect\ref{prop}) and the circled crosses denote the poles of (%
\protect\ref{peps})}
\label{Fig1}
\end{figure}

Multiply (\ref{prop}) by $\pm $ and, following \cite{LWgrav}, write the
outcome as 
\begin{equation*}
\pm \frac{p^{2}-m^{2}}{(p^{2}-m^{2})^{2}}.
\end{equation*}%
Then, eliminate the singularity by introducing an infinitesimal width $%
\mathcal{E}$ as follows, to define the new propagators 
\begin{equation}
\mathbb{G}_{\pm }(p,m,\mathcal{E}^{2})=\pm \frac{p^{2}-m^{2}}{%
(p^{2}-m^{2})^{2}+\mathcal{E}^{4}}=\pm \frac{1}{2}\left[ G_{+}(p,m,\mathcal{E%
}^{2})-G_{-}(p,m,\mathcal{E}^{2})\right] .  \label{peps}
\end{equation}%
A first reason why $\mathbb{G}_{\pm }(p,m,\mathcal{E}^{2})$ does not truly
propagate a particle is that it vanishes on shell at $\mathcal{E}>0$.
Alternatively, the relative sign between $G_{+}$ and $G_{-}$ in the last
expression suggests that $\mathbb{G}_{\pm }$ propagates a sort of
\textquotedblleft siamese particle/ghost pair\textquotedblright\ and that
the particle and the ghost somewhat compensate each other. This is what
gives the fake particle.

Doing (\ref{peps}), we basically split the poles of (\ref{prop}) into pairs
of complex conjugates poles, as shown in fig. \ref{Fig1}. The construction
and the Wick rotation of the Euclidean theory suggest that, when the
propagator $\mathbb{G}_{\pm }$ is used inside the Feynman diagrams, the loop
energy $p^{0}$ must be integrated along the path shown in the same fig. \ref%
{Fig1}, which passes under the left pair of complex conjugate poles and over
the right pair. This is called \textit{Lee-Wick (LW)\ integration
prescription}, because it first appeared in the Lee-Wick models \cite%
{leewick}. We comment on the relation between fakeons and Lee-Wick models
below.

There exist three types of fakeons. The fakeon with propagator $\mathbb{G}%
_{+}(p,m,\mathcal{E}^{2})$ is called \textit{fakeon plus}. The fakeon with
propagator $\mathbb{G}_{-}(p,m,\mathcal{E}^{2})$ is called \textit{fakeon
minus}. For the moment, we ignore the third type of fakeon, to be described
later on.

\begin{figure}[t]
\begin{center}
\includegraphics[width=8truecm]{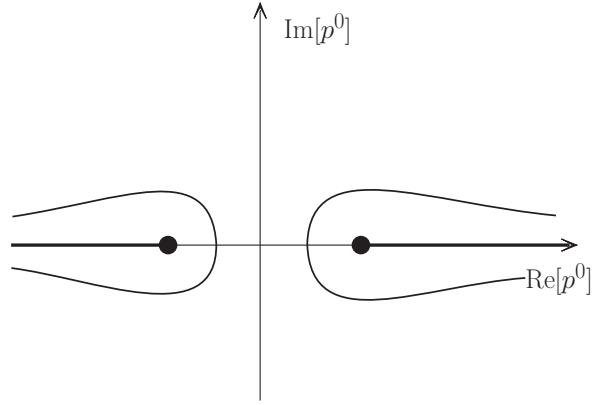}
\end{center}
\caption{Analyticity fails in extended regions instead of branch cuts}
\label{regions}
\end{figure}

To make the arguments more explicit, we consider the bubble diagram as an
example. Integrating the loop energy $k^{0}$ along the Lee-Wick integration
path and the loop space momentum $\mathbf{k}$ on $\mathbb{R}^{3}$, we get 
\begin{equation}
i\mathcal{M}(p)=c\int_{\mathbb{LW}}\frac{\mathrm{d}k^{0}}{2\pi }\int_{%
\mathbb{R}^{3}}\frac{\mathrm{d}^{3}\mathbf{k}}{(2\pi )^{3}}\mathbb{G}_{\pm
}(p-k,m_{1},\mathcal{E}^{2})\mathbb{G}_{\pm }(k,m_{2},\mathcal{E}^{2})=\int_{%
\mathbb{R}^{3}}\frac{d^{3}\mathbf{k}}{(2\pi )^{3}}w(p,\mathbf{k})
\label{bubble}
\end{equation}%
for some integrand $w(p,\mathbf{k})$, where $c$ collects the coupling
constants and the combinatorial factor. For the arguments that follow, it
does not matter whether the propagators are both $\mathbb{G}_{+}$, or both $%
\mathbb{G}_{-}$, or one and one.

The integral is always regular. When $p$ is such that $w(p,\mathbf{k})$ is
regular for all $\mathbf{k}$, the function $\mathcal{M}(p)$ is analytic in $p
$. Moreover, when $p$ is real and $\mathcal{M}(p)$ is analytic in $p$, $%
\mathcal{M}(p)$ is real. Analyticity has potential problems when $w(p,%
\mathbf{k})$ is singular. The $w(p,\mathbf{k})$ singularities originate the
discontinuity of $\mathcal{M}$, which is also the imaginary part $\mathrm{Im}%
\mathcal{M}$. By the optical theorem, specifically the identity (\ref{cutdd}%
), the imaginary part gives the total cross section of a scattering process,
where the external particle of momentum $p$ decays into the two particles
circulating in the loop, of momenta $p-k$ and $k$. The conditions for having
such a process read 
\begin{equation}
|p^{0}|=\sqrt{(\mathbf{p}-\mathbf{k})^{2}+m_{1}^{2}\pm i\mathcal{E}^{2}}+%
\sqrt{\mathbf{k}^{2}+m_{2}^{2}\pm i\mathcal{E}^{2}}  \label{condition}
\end{equation}%
(all four possibilities occurring) \cite{LWunitarity}. Note that both
frequencies on the right-hand side come with positive signs. The other
possibilities are excluded by the LW integration prescription.

In the limit $\mathcal{E}\rightarrow 0$, analyticity fails in branch cuts,
whose branch points are the thresholds of the process. Instead, when $%
\mathcal{E}\neq 0$ the branch cuts are replaced by extended regions $%
\mathcal{\tilde{A}}$, obtained by plotting (\ref{condition}) for $\mathbf{k}%
\in\mathbb{R}^{3}$ with $\mathbf{p}$ fixed. Examples of such regions are
shown in fig. \ref{regions}. Inside those regions the result of the integral
(\ref{bubble}) is not analytic and not Lorentz invariant.

Since all that matters is the limit $\mathcal{E}\rightarrow 0$, one might
wonder why we pay attention to the properties of the integral at $\mathcal{E}%
>0$. The reason is that if we take the limit too quickly we miss the new
quantization prescription. Indeed, if we work at $\mathcal{E}=0$, when the
regions $\mathcal{\tilde{A}}$ shrink to branch cuts, we can circumvent the,
say, right branch point by coming from above or from below, i.e. from the
upper or lower side of the complex plane (check the left picture of fig. \ref%
{Fig4}). The two options correspond to the Feynman prescription and its
conjugate, respectively, which give nothing new. It is like replacing the
propagators of (\ref{bubble}) with two $G_{+}$ or two $G_{-}$, respectively:
we have no coexistence of $G_{+}$ and $G_{-}$ in the same loop integral.

\begin{figure}[t]
\begin{center}
\includegraphics[width=16truecm]{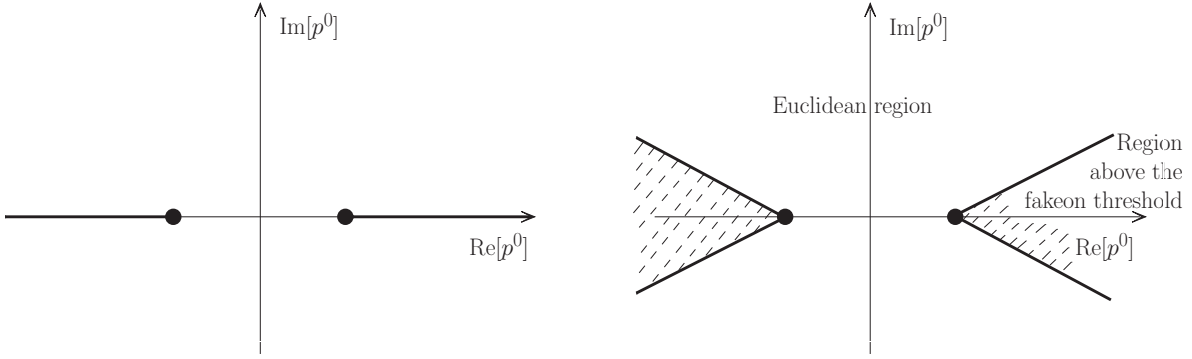}
\end{center}
\caption{Analyticity versus regionwise analyticity}
\label{Fig4}
\end{figure}

Before taking the limit $\mathcal{E}\rightarrow 0$ we have a new
possibility, which turns out to be the only way to mix $G_{\pm }$ in the
same loop integral consistently with unitarity: we go inside a region $%
\mathcal{\tilde{A}}$ (i.e. we choose a $p^{0}$ that belongs to the portion
of the real axis that is contained in the region), evaluate the integral
there and \textit{then} take the limit $\mathcal{E}\rightarrow 0$. In such a
limit, analyticity and Lorentz invariance are recovered.

Moreover, the discontinuity $\text{Im}\mathcal{M}$ of $\mathcal{M}$
disappears, because the operation is symmetric under reflections with
respect to the real axis, so it cannot generate an imaginary part. Thanks to
this, the fakeon can be projected away from the physical spectrum $V$.
Actually, this is the only option we have, if we want unitarity (i.e. the
optical theorem). The candidate physical process becomes a fake, with zero
probability to occur, and is thrown away as well. Precisely, the identities (%
\ref{cutdd}) are satisfied as follows: the left-hand side and the middle
expression vanish thanks to the fakeon prescription, while the right-hand
side vanishes thanks to the fakeon projection, since fakeons are excluded
from the set of physical asymptotic states. This is how the fakeon
prescription/projection turns out to be compatible with unitarity. Instead,
no projection can be consistent with the Feynman prescription, since the
right-hand side of (\ref{cutd}) would vanish by definition, but the
left-hand side would continue to be nontrivial.

A mathematical theorem \cite{LWformulation,fakeons} ensures that the result
of the $\mathcal{M}$ evaluation inside the region $\mathcal{\tilde{A}}$ is
equal to the arithmetic average of the two analytic continuations that
circumvent the branch point. This operation is called \textit{average
continuation}.

These properties generalize to all diagrams \cite{fakeons}. In conclusion,
we can have the propagators $G_{+}$ and $G_{-}$ coexist in the form $\mathbb{%
G}_{\pm }$, provided we treat them as just explained. Their poles do not
correspond to physical particles, nor ghosts, but fake particles, or
fakeons. The scattering processes involving at least one fakeon are fake
processes and their thresholds must be circumvented by means of the average
continuation. Then, their probability to occur vanishes, which makes it
possible to project them away and have unitarity.

The average continuation is an unambiguous, nonanalytic operation to
circumvent branch points. It associates an analytic function $f_{\text{AV}%
}(z)$ to an analytic function $f(z)$. However, in general, $f_{\text{AV}}(z)$
is not analytically related to $f(z)$. As a consequence, the hyperplane $%
\mathcal{P}$ of the complexified external momenta turns out to be divided
into disjoint regions of analyticity and the amplitudes $\mathcal{M}$ are
separately analytic in each region. We call this property \textit{regionwise
analyticity}.

The main region is the Euclidean one, which contains the purely imaginary
energies. There, the Wick rotation is analytic, because no average
continuation is necessary. The other regions can be reached unambiguously
from the main one by means of the average continuation. It is worth to
emphasize that the relative simplicity of the average continuation makes
calculations doable with not much more effort than usual.

If the theory contains only physical particles, then we have analyticity,
which means that it is sufficient to compute a loop diagram in \textit{any}
open subset of $\mathcal{P}$ to derive it everywhere in $\mathcal{P}$ by
means of the analytic continuation. If the theory contains fakeons, then we
have regionwise analyticity. In particular, it is sufficient to compute a
loop diagram in any open subset of the Euclidean region to derive it
everywhere in $\mathcal{P}$ by means of the average continuation. It is not
sufficient to know the amplitude in regions different from the Euclidean
one. Indeed, there are many functions $f(z)$ whose average continuation $f_{%
\text{AV}}(z)$ vanishes identically (for example, $f(z)=\sqrt{z}$), so in
general it is impossible to reconstruct an $f(z)$ from its average
continuation $f_{\text{AV}}(z)$.

In fig. \ref{Fig4} analyticity and regionwise analyticity are compared. The
basic difference is that some branch cuts are replaced by extended regions.

We have anticipated that there exists a third type of fakeon, which is the 
\textit{thick fakeon}. The fakeon plus and the fakeon minus have
infinitesimal widths $\mathcal{E}$ and the limit $\mathcal{E}\rightarrow 0$
must be taken as explained above. Instead, the thick fakeon has a finite
nonzero width $\mathcal{E}$ and propagator 
\begin{equation}
\mathbb{G}_{\text{c}}(p,m,\mathcal{E}^{2})=aG_{+}(p,m,\mathcal{E}%
^{2})-a^{\ast }G_{-}(p,m,\mathcal{E}^{2}),  \label{thick}
\end{equation}%
where $a$ is a complex coefficient. The squared masses $m^{2}\pm i\mathcal{E}%
^{2}$ are also complex, with a nonnegative real part $m^{2}$. There is no
option to quantize the poles of $\mathbb{G}_{\text{c}}$ as physical
particles.

The thick fakeons are also called Lee-Wick fakeons, since they are the ones
that appear in the Lee-Wick models \cite{leewick}. Before moving on, we
recall how the quantization of the thick fakeon works, because it requires
supplementary operations. Let us start over, from the Lee-Wick prescription\
of fig. \ref{Fig1} for the integrals on the loop energies. That prescription
is not sufficient to define the Lee-Wick models properly, because it leads
to violations of Lorentz invariance and ambiguities \cite{nakanishi,cutkosky}%
. Extra prescriptions were proposed right after the papers of Lee and Wick 
\cite{cutkosky}, but\ they did not remove the ambiguities completely.

Since the Lee-Wick integration prescription involves complex values of the
loop energies, it is not possible to have Lorentz invariance above the
fakeon thresholds at finite $\mathcal{E}$ by integrating on real loop space
momenta. It is necessary to deform the integration domain on the loop space
momenta to include complex values for them as well \cite%
{LWformulation,fakeons}. It can be shown that it is possible to arrange the
deformation so as to squeeze the extended regions $\mathcal{\tilde{A}}$ onto
branch cuts even if $\mathcal{E}>0$. At the end, interestingly enough, the
amplitude above the fakeon threshold is still given by the average
continuation. Again, the scattering processes that involve the fakeons have
zero probability to occur and can be projected away, as required by
unitarity.

The Lee-Wick\ prescription and the deformations of the integration domains
on the loop space momenta are the operations that define the nonanalytic
Wick rotation of the Euclidean theory, which is equivalent to the average
continuation. What makes the whole construction work, by removing the
difficulties that prevented to find a sound definition of the Lee-Wick
models decades ago, is the concept of fake particle.

A weakness of the Lee-Wick models is that they are super-renormalizable.
Since there are infinitely many of them, we have no way to decide which is
the right one. Having infinitely many theories with finitely many parameters
is not so different from having one nonrenormalizable theory with infinitely
many parameters, like Einstein gravity, after it is equipped with all the
counterterms generated by renormalization. Moreover, nature does not seem to
favor super-renormalizable theories for high-energy physics.

Instead, the theory of quantum gravity (\ref{SQG})-(\ref{SQG2}) considered
here, defined using the fakeon prescription (\ref{peps}) for the spin-2
massive field $\chi _{\mu \nu }$, is essentially unique, because it is
strictly renormalizable. More precisely, it contains a finite number of
independent parameters and can be quantized in a finite number of consistent
ways. Under many respects, it is the theory that is most similar to the
standard model, to which the matter sector can be attached with no effort.

\subsection{Prescription and projection}

\label{presco}

Summarizing, the quantization\textit{\ }prescription is defined by
introducing two infinitesimal widths $\epsilon $ and $\mathcal{E}$ in the
propagators\ as follows:

($a$) replace $p^{2}$ with $p^{2}+i\epsilon $ everywhere in the
denominators, where $p$ denotes the momentum;

($b$) treat the poles you want to convert into fakeons by means of
replacements of the form%
\begin{equation}
\frac{1}{p^{2}-m^{2}+i\epsilon }\rightarrow \frac{p^{2}-m^{2}}{%
(p^{2}-m^{2}+i\epsilon )^{2}+\mathcal{E}^{4}};  \label{nopri}
\end{equation}

($c$) calculate the diagrams in the Euclidean framework, nonanalytically
Wick rotate them as explained above, then make $\epsilon $ tend to zero
first and $\mathcal{E}$ tend to zero last.

Clearly, these rules are meant to be applied in momentum space. It is harder
to work out the quantization rules directly in coordinate space and study
the nonanalytic Wick rotation there. Thus, we might also want to specify that

($d$) the amplitudes and the loop integrals must be evaluated in momentum
space and then Fourier transformed to coordinate space.

An equivalent set of quantization rules is obtained by combining ($a$) with (%
$d$) and the requirement that, in evaluating the loop integrals,

($e$) every threshold involving a fakeon must be overcome by means of the
average continuation.

Now, let us quantize the actions (\ref{SQG}) and (\ref{SQG2}). Formula (\ref%
{ss}) shows that the $\chi _{\mu \nu }$ quadratic action has the wrong
overall sign, so the field $\chi _{\mu \nu }$ must be quantized as a fakeon,
according to step ($b$). Instead, $\phi $ can be quantized either as a
fakeon or a physical particle. Depending on which option we choose, we have
a graviton/fakeon/fakeon (GFF) theory or a graviton/scalar/fakeon (GSF)
theory.

By the optical theorem, the processes that involve the fakeons have zero
probabilities to occur. Then, to have consistency we must project away the
fakeons from the physical spectrum. So doing, the theory is unitary (and
renormalizable).

The physical states are obtained by acting on the vacuum $|0\rangle $ by
means of the creation operators of the physical particles only, ignoring the
creation operators of the fakeons (which are $a_{\chi }^{\dag }$ and $%
a_{\phi }^{\dag }$ in the GFF theory, just $a_{\chi }^{\dag }$ in the GSF\
theory). This defines, after Cauchy completion, the Fock space $V$ of the
physical states, which is a proper subspace of the total Fock space $W$. The
projection $W\rightarrow V$ is called \textit{fakeon projection}. The free
Hamiltonian is bounded from below in $V$ (but obviously not in $W$).

It is helpful to make a comparison between the fakeon projection and the 
\textit{gauge projection}, by which we mean the projection involved in the
gauge theories, which concerns the Faddeev-Popov ghosts and the longitudinal
and temporal components of the gauge fields. Consider a gauge-fixed action $%
S_{\text{gf}}$. Usually, the gauge-fixing condition is not solved
explicitly, because if we inserted its solution into $S_{\text{gf}}$ we
would obtain a nonlocal action, which is much more difficult to deal with.
It is preferable to keep the Faddeev-Popov ghosts and the longitudinal and
temporal modes of the gauge fields till the very end, work in a local
framework and perform the gauge projection only when strictly needed.

The fakeon projection also introduces nonlocalities (see sections \ref{toy}
and \ref{class} for details). The virtue of the interim, unprojected actions
(\ref{SQG}) and (\ref{SQG2}) is that they allow us to work within local
frameworks, pretty much like the gauge-fixed actions. However, there is a
crucial difference between the fakeon projection and the gauge projection.
By changing the gauge fixing it is possible to reach a gauge (the Coulomb
gauge), where the gauge projection acts not only on the initial and final
states, but even inside the loop diagrams. Thanks to this, the gauge modes
disappear from everywhere. It is not possible to achieve an analogous result
by means of the fakeon projection, which does act on the initial and final
states, but cannot reach inside the loop diagrams. The net result is that\
the fake particles leave an important remnant, which is the violation of
causality at energies larger than their masses. This is also the reason, why 
\textit{the fakeons must be massive}, otherwise causality would be violated
at all energies. We stress that the fakeon projection is the only projection
known at present that is consistent with unitarity even it does not follow
from a gauge or symmetry principle.

\section{Microcausality}

\setcounter{equation}{0}\label{predi}

Thanks to the average continuation, calculating loop diagrams with the
fakeon prescription does not require much more effort than calculating
diagrams with the ordinary prescriptions \cite{UVQG,absograv}. Among the
first things to compute, we mention the one-loop self-energy diagrams, which
give the physical masses $\bar{m}$ and the physical widths $\Gamma $.

In the case of the fakeons, if we resum the bubble diagrams $B$, we get the
dressed propagators 
\begin{equation}
\mathbb{\bar{G}}_{\pm }=\mathbb{G}_{\pm }+\mathbb{G}_{\pm }B\mathbb{G}_{\pm
}+\mathbb{G}_{\pm }B\mathbb{G}_{\pm }B\mathbb{G}_{\pm }+\cdots =\frac{1}{%
\mathbb{G}_{\pm }^{-1}-B}.  \label{resum}
\end{equation}%
After the resummation, we can take $\mathcal{E}$ to zero, which gives,
around the physical peak $p^{2}=\bar{m}^{2}$,%
\begin{equation*}
\mathbb{\bar{G}}_{\pm }\sim \pm \frac{Z}{p^{2}-\bar{m}^{2}+i\bar{m}\Gamma
_{\pm }}=\pm ZG_{+}(p,\bar{m},\bar{m}\Gamma _{\pm }),
\end{equation*}%
where $Z$ is the normalization factor. The optical theorem implies 
\begin{equation*}
\mathrm{Im}[\mp ZG_{+}(p,\bar{m},\bar{m}\Gamma _{\pm })]=\frac{\bar{m}Z(\pm
\Gamma _{\pm })}{(p^{2}-\bar{m}^{2})^{2}+\bar{m}^{2}\Gamma _{\pm }^{2}}%
\geqslant 0,
\end{equation*}%
i.e. $\Gamma _{+}>0$, $\Gamma _{-}<0$. We thus learn that a fakeon plus has
a positive width, while a fakeon minus has a negative width. Moreover, the
limits $\Gamma _{\pm }\rightarrow 0^{\pm }$ give%
\begin{equation}
\lim_{\Gamma _{\pm }\rightarrow 0^{\pm }}\mathrm{Im}[\mp ZG_{+}(p,\bar{m},%
\bar{m}\Gamma _{\pm })]\sim \pi Z\delta (p^{2}-\bar{m}^{2})  \label{imma}
\end{equation}%
in both cases. This result shows that if we just watch the decay products of
a fakeon, we have the illusion that a true particle exists, no matter how
small its width is and no matter whether the width is positive or negative.

At the same time, the resummation (\ref{resum}) is legitimate only if $%
p^{2}-m^{2}$ is large enough, which means that it misses the contact terms $%
\delta (p^{2}-m^{2})$, $\delta ^{\prime }(p^{2}-m^{2})$, etc. In total, the
contact terms are 
\begin{equation}
\sigma \pi Z\delta (p^{2}-\bar{m}^{2})  \label{immac}
\end{equation}%
for $\Gamma _{\pm }\rightarrow 0^{\pm }$, where $\sigma =1,0,-1$ in the case
of a physical particle, a fakeon and a ghost, respectively. For example, at
the tree level a physical particle gives (\ref{opti}), a ghost gives the
opposite of (\ref{opti}), while a fakeon gives 
\begin{equation}
\mathrm{Im}[-\mathbb{G}_{\pm }(p,m,\mathcal{E}^{2})]=\mathrm{Im}[-\mathbb{G}%
_{\text{c}}(p,m,\mathcal{E}^{2})]=0.  \label{optio}
\end{equation}%
This formula tells us what we see if we do not let the fakeons decay, but
try to detect them \textquotedblleft on the fly\textquotedblright . The
answer is that we see precisely nothing. Only in the case of a physical
particle what we infer from the indirect observation, which is encoded in
formula (\ref{imma}), coincides with what we get from the direct
observation, which is given by formula (\ref{immac}). 
\begin{figure}[t]
\begin{center}
\includegraphics[width=8truecm]{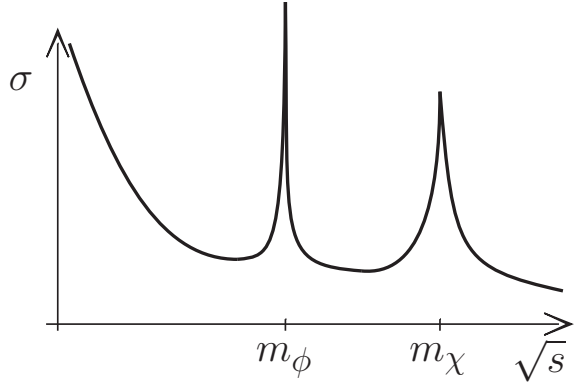}
\end{center}
\caption{Cross section}
\label{crosssec}
\end{figure}

In other words, if we restrict our attention to the processes where the
fakeon does decay, the total cross section gives a plot like the one shown
in fig. \ref{crosssec}, which coincides with the plot we would see in the
case of a physical particle. However, a physical particle can also be
detected before it decays (at least in principle), when it is still
\textquotedblleft alive\textquotedblright , since it belongs to the physical
spectrum. On the other hand, a fakeon cannot be detected directly, because
the probability to produce it is zero. The only ways to \textquotedblleft
see\textquotedblright\ a fakeon are indirect, by means of its decay products
and the interactions it mediates. These facts justify the name,
\textquotedblleft fake particle\textquotedblright , or fakeon, for the new
entity.

Since $\chi _{\mu \nu }$ is a fakeon minus, its width $\Gamma _{\chi }$ is
negative. Precisely, the calculation gives \cite{absograv} 
\begin{equation}
\Gamma _{\chi }=-C\frac{m_{\chi }^{3}}{M_{\mathrm{Pl}}^{2}},\qquad C=\frac{1%
}{120}(N_{s}+6N_{f}+12N_{v}),  \label{gamma}
\end{equation}%
in the case of the GFF theory, where $N_{s}$, $N_{f}$ and $N_{v}$ are the
numbers of (physical) scalars, Dirac fermions (plus one half the number of
Weyl fermions) and gauge vectors, respectively. We are assuming that the
masses of the matter fields are much smaller than $m_{\chi }$, otherwise
there are corrections \cite{absograv}. In the case of the GSF theory, there
is also a correction due to $\phi $, which depends on $m_{\phi }$. The
graviton and the fakeons do not contribute to $\Gamma _{\chi }$.

The negative sign of $\Gamma _{\chi }$ signals the violation of
microcausality at the quantum level. Consider the Breit-Wigner distribution
and its Fourier transform:%
\begin{equation}
\frac{i}{E-\bar{m}+i\frac{\Gamma }{2}},\qquad G_{\mathrm{BW}}(t)=\mathrm{sgn}%
(t)\hspace{0.01in}\theta (\Gamma t)\exp \left( -i\bar{m}t-\frac{\Gamma t}{2}%
\right) ,  \label{Gt}
\end{equation}%
where $\mathrm{sgn}(t)$ is the sign of $t$. Observe that $\exp \left(
-\Gamma t/2\right) $ is always a dumping factor. If $J(t)$ denotes an
external source, the response reads 
\begin{equation}
\int_{-\infty }^{+\infty }\mathrm{d}t^{\prime }G_{\mathrm{BW}}(t-t^{\prime
})J(t^{\prime })=-\int_{t}^{\infty }\mathrm{d}t^{\prime }\hspace{0.01in}%
\mathrm{e}^{-\left( i\bar{m}+\frac{\Gamma }{2}\right) (t-t^{\prime
})}J(t^{\prime })  \label{response}
\end{equation}%
for $\Gamma <0$. This formula shows that when the width is negative it is
necessary to know the values of the source $J$ in the future. However, due
to the (dumping or oscillating) exponential factors that appear on the
right-hand side of (\ref{response}), it is sufficient, at the practical
level, to anticipate the source $J(t^{\prime })$ just for a little bit of
future, such that $t^{\prime }-t\lesssim \tau \equiv \min (1/\bar{m}%
,2/|\Gamma |)$.

In other words, time, as well as past, present and future, and the concepts
of cause and effect, lose meaning for intervals smaller than $\tau $.
However, as long as $\tau $ is short enough, the possibility of having
violations of microcausality in nature is compatible with experiments.

Formula (\ref{gamma}) shows that the violation of causality cannot be
eliminated, since adding physical matter fields can at most increase $%
|\Gamma _{\chi }|$, while adding fakeons leaves $|\Gamma _{\chi }|$
invariant.

As said, the massive scalar $\phi $ can be a physical particle or a fakeon
plus. In either case, its width is positive.

We have no experimental or logical reason to claim that causality should
hold up to infinite energies or zero relative distances. On the contrary, we
view the violation of microcausality as a major prediction of quantum
gravity and concentrate on finding ways to detect its effects. The fakeon
mass $m_{\chi }$ is a free parameter at the moment. Its actual value might
be smaller, or even much smaller, than the Planck mass $M_{\mathrm{Pl}}$.
Thanks to this, it might be possible to detect the first signs of quantum
gravity various orders of magnitude below the Planck scale. Moreover, as we
show in the next sections, we might be able to study the problem
noperturbatively, by working out the corrections to the field equations of
general relativity.

To make progress in this direction, it is helpful to study the remnants of
the fakeons in the classical limit. At the tree level, the steps ($c$) and ($%
e$) can be skipped, so the fakeon has the free propagator%
\begin{equation}
\frac{p^{2}-m^{2}}{(p^{2}-m^{2})^{2}+\mathcal{E}^{4}}=\frac{1}{2}\left[ 
\frac{1}{(p^{0}+i\epsilon )^{2}-\mathbf{p}^{2}-m^{2}}+\frac{1}{%
(p^{0}-i\epsilon )^{2}-\mathbf{p}^{2}-m^{2}}\right] =\mathcal{P}\frac{1}{%
p^{2}-m^{2}}.  \label{fprc}
\end{equation}%
The equalities are meant in the sense of distributions, for $\epsilon
\rightarrow 0$ and $\mathcal{E}\rightarrow 0$. We obtain the Cauchy
principal value, which is also the half sum of the retarded and advanced
Yukawa potentials. Here, the violation of microcausality is due to the
advanced Green function.

Since different quantization prescriptions may have the same classical
limit, it is not possible to infer the fakeon prescription from the
principal value (\ref{fprc}). Actually, as it stands, (\ref{fprc}) leads to
wrong results, if applied to the loop integrals, because it misses steps ($c$%
) and ($e$), hence it generates the problems of ref. \cite{ugo} and violates
the optical theorem. We stress again that the loop integrals must be
calculated from their Euclidean versions by performing the nonanalytic Wick
rotation of ref.s \cite{LWformulation,fakeons}, or crossing the fakeon
thresholds inside the amplitudes by means of the average continuation.

\section{Classicization: a toy model}

\setcounter{equation}{0}\label{toy}

The next goal is to analyze the effects of the fakeons on causality in the
classical limit. In this section, we study the nonrelativistic particle as a
toy model to illustrate the main properties. Consider the higher-derivative
Lagrangian%
\begin{equation}
\mathcal{L}_{\text{HD}}=\frac{m}{2}(v^{2}-\tau ^{2}a^{2})-V(x,t),
\label{lagen2}
\end{equation}%
where $x$ is the coordinate ($v=\dot{x}$, $a=\ddot{x}$) and $\tau $ is a
real constant. To begin with, let us take $V(x,t)=-xF_{\text{ext}}(t)$,
where $F_{\text{ext}}(t)$ is an external force. The equation of motion $%
m(a+\tau ^{2}\ddot{a})=F_{\text{ext}}$ has four independent solutions,
unless we restrict the configuration space. The restriction can be achieved
by writing 
\begin{equation}
ma=\frac{1}{1+\tau ^{2}\frac{d^{2}}{dt^{2}}}F_{\text{ext}}\equiv \tilde{G}F_{%
\text{ext}}  \label{preq}
\end{equation}%
and giving a prescription for the Green function $\tilde{G}$ that acts on $%
F_{\text{ext}}$. The distribution $\tilde{G}$ that follows from the
classical\ fakeon prescription (\ref{fprc}) is%
\begin{equation}
G_{F}(u,\tau )=\frac{\sin (|u|/\tau )}{2\tau },  \label{fakeonG}
\end{equation}%
where the subscript $F$ stands for \textquotedblleft
fakeon\textquotedblright . The projected equation of motion is then%
\begin{equation}
ma(t)=\int_{-\infty }^{\infty }\mathrm{d}u\hspace{0in}\hspace{0.01in}%
G_{F}(u,\tau )F_{\text{ext}}(t-u)\equiv \left\langle F_{\text{ext}%
}(t)\right\rangle  \label{average}
\end{equation}%
and its degrees of freedom are just the initial position and the initial
velocity of the particle. Since%
\begin{equation}
\lim_{\tau \rightarrow 0}G_{F}(u,\tau )=\delta (u),  \label{limit}
\end{equation}%
the equation becomes $ma=F_{\text{ext}}$ in the limit $\tau \rightarrow 0$,
as expected\footnote{%
The quickest way to prove (\ref{limit}), pointed out to us by L. Bracci, is
to take the derivative of the distribution $\mathrm{sgn}(u)\cos (u/\tau )$,
which tends to zero by the Riemann-Lebesgue theorem.}.

The fakeon Green function (\ref{fakeonG}) makes the average $\left\langle F_{%
\text{ext}}(t)\right\rangle $ sensitive to both the past and the future,
which means that the fakeon solutions disappear at the expenses of
microcausality. \ However, the violation of microcausality is
\textquotedblleft small\textquotedblright , since its effects are averaged
away by the oscillating behavior of $G_{F}$ for $|u|\gg \tau $. As in the
previous section, it is sufficient to know just \textquotedblleft a little
bit of future\textquotedblright , say $|u|\lesssim \tau $, to predict the
future.

If we introduce an auxiliary coordinate $Q$ and make the redefinition $x=q+Q$%
, we obtain the equivalent Lagrangian%
\begin{equation}
\mathcal{L}(q,Q,t)=\frac{m}{2}\dot{q}^{2}-\frac{m}{2}\dot{Q}^{2}+\frac{m}{%
2\tau ^{2}}Q^{2}-V(q+Q,t).  \label{lqQ}
\end{equation}%
The $Q$-quadratic part has the wrong sign, as expected. The equations of
motion%
\begin{equation}
m\ddot{q}=-\frac{\partial V}{\partial q}(q+Q,t),\qquad m\ddot{Q}+\frac{m}{%
\tau ^{2}}Q=\frac{\partial V}{\partial q}(q+Q,t),  \label{eomqQ}
\end{equation}%
can be projected by interpreting $Q$ as a fake coordinate. We assume that
the potential $V$ can be treated perturbatively and solve the $Q$ equation
of motion by means of the classical fakeon prescription, which gives $%
Q(q,t)=-\tau ^{2}\langle \ddot{q}\rangle $. Then we substitute the result
back into (\ref{eomqQ}). So doing, we obtain the projected equation of
motion for $q$:%
\begin{equation}
m\ddot{q}=-\left. \frac{\partial V(x,t)}{\partial x}\right\vert _{x=\langle
q\rangle },  \label{eomq}
\end{equation}%
where the average $\langle q(t)\rangle $ is defined by the last equality of
eq. (\ref{average}) with $F_{\text{ext}}\rightarrow q$.

We stress that, since the fakeon prescription comes from quantum field
theory, which is formulated perturbatively, the projected classical
equations are also understood perturbatively. The fakeon solutions of (\ref%
{eomqQ}) drop out, because they are not perturbative solutions of (\ref{eomq}%
). If we want to treat $V$ nonperturbatively, we need to resum the expansion.

The projected Lagrangian $\mathcal{L}_{r}(q,t)$ can be obtained by inserting
the solution $Q(q,t)=-\tau ^{2}\langle \ddot{q}\hspace{0.01in}\rangle $ back
into (\ref{lqQ}), which gives%
\begin{equation}
\mathcal{L}_{r}(q,t)=\mathcal{L}(q,Q(q,t),t)=\frac{m}{2}\left( \langle \dot{q%
}\rangle ^{2}+2\tau ^{2}\langle \dot{q}\rangle \langle \dddot{q}\hspace{%
0.01in}\hspace{0.01in}\rangle +\tau ^{2}\langle \ddot{q}\hspace{0.01in}%
\rangle ^{2}\right) -V(\langle q\rangle ,t).  \label{Lr}
\end{equation}%
Its Lagrange equations are indeed the projected equations of motion (\ref%
{eomq}).

Given an arbitrary Lagrangian $L(q,\dot{q},\ddot{q},\dddot{q},\cdots ,t)$,
the energy is%
\begin{equation}
E=-L+\dot{q}\sum_{n=1}^{\infty }\frac{\overleftarrow{\partial _{t}}%
^{n}-(-1)^{n}\overrightarrow{\partial _{t}}^{n}}{\overleftarrow{\partial _{t}%
}+\overrightarrow{\partial _{t}}}\frac{\partial L}{\partial q^{(n)}},
\label{energy}
\end{equation}%
where $\partial _{t}=d/dt$, the arrows specify whether the derivatives act
to the left or the right, and $q^{(n)}=\partial _{t}^{n}q$. The ratio of
derivative operators appearing in (\ref{energy}) must be simplified by means
of the polynomial identity $(x^{n}-(-1)^{n}y^{n})/(x+y)=x^{n-1}+\cdots
-(-1)^{n}y^{n-1}$. It is easy to check that the Lagrange equations imply $%
dE/dt=-\partial L/\partial t$, so $E$ is conserved if $t$ is a cyclic
coordinate. Moreover, under arbitrary changes of variables $u=u(q,\dot{q},%
\ddot{q},\dddot{q},\cdots )$ that do not depend explicitly on $t$, $E$ can
at most go into a function of $E$, so we can use formula (\ref{energy}) with
the variables we want. To apply it to $L=\mathcal{L}_{r}$ it is convenient
to use the variable $u=\langle q\rangle =x$, which easily leads to the
expression%
\begin{equation}
E^{\prime }=\frac{m}{2}\left( \langle \dot{q}\rangle ^{2}+2\tau ^{2}\langle 
\dot{q}\rangle \langle \dddot{q}\hspace{0.01in}\hspace{0.01in}\rangle -\tau
^{2}\langle \ddot{q}\hspace{0.01in}\rangle ^{2}\right) +V(\langle q\rangle
,t).  \label{ener}
\end{equation}

An as example, consider the harmonic oscillator, $V=m\omega ^{2}x^{2}/2$.
The equations of motion give%
\begin{equation}
(\nu ^{2}-\omega ^{2}-\tau ^{2}\nu ^{4})\tilde{x}(\nu )=0,  \label{equa}
\end{equation}%
where $\tilde{x}(\nu )$ is the Fourier transform of $x(t)$. If $\omega
>1/(2\tau )$ the polynomial $\nu ^{2}-\omega ^{2}-\tau ^{2}\nu ^{4}$ has two
pairs of complex conjugate zeros, which correspond to thick fakeons. Thus,
the projected subspace of configurations is empty. Instead, for $\omega
\leqslant 1/(2\tau )$ the polynomial has four real zeros, but only two of
them give solutions that have regular $\tau \rightarrow 0$ limits. They read%
\begin{equation}
x(t)=A\cos (\Omega t+\varphi _{0}),\qquad \text{where }\Omega =\frac{1}{\tau 
\sqrt{2}}\sqrt{1-\sqrt{1-4\tau ^{2}\omega ^{2}}}  \label{solos}
\end{equation}%
and $A$, $\varphi _{0}$ are constants. The other two solutions are fakeons
minus, as can be seen from the residues of their propagators, obtained by
replacing the right-hand side of (\ref{equa}) with 1. Projecting the fakeons
away, the final result is a harmonic oscillator with the modified frequency $%
\Omega $. The energy (\ref{ener}) is%
\begin{equation}
E^{\prime }=\frac{m}{2}\left( \dot{x}^{2}+2\tau ^{2}\dot{x}\dddot{x}-\tau
^{2}\ddot{x}^{2}\right) +V(x)=\frac{m}{2}A^{2}\Omega ^{2}\sqrt{1-4\tau
^{2}\omega ^{2}}  \label{eosc}
\end{equation}%
and is positive definite on the solutions (\ref{solos}).

To quantize the theory (\ref{lqQ}) in the case $\omega \leqslant 1/(2\tau )$%
, we define the annihilation operators%
\begin{equation*}
a_{\xi }=\sqrt{\frac{m\Omega }{2}}\left( \xi +\frac{i}{m\Omega }P_{\xi
}\right) ,\qquad a_{\eta }=\sqrt{\frac{m\tilde{\Omega}}{2}}\left( \eta +%
\frac{i}{m\tilde{\Omega}}P_{\eta }\right) ,
\end{equation*}%
where $\tilde{\Omega}=\sqrt{1-\tau ^{2}\Omega ^{2}}/\tau $, $\xi =q\cosh
\theta -Q\sinh \theta $, $\eta =Q\cosh \theta -q\sinh \theta $, $P_{\xi
}=p\cosh \theta +P\sinh \theta $, $P_{\eta }=P\cosh \theta +p\sinh \theta $
and $\theta $ is such that $\tanh \theta =\Omega ^{2}/\tilde{\Omega}^{2}$.
The commutation rules $[p,q]=[P,Q]=-i$ lead to $[a_{\xi },a_{\xi }^{\dag
}]=[a_{\eta },a_{\eta }^{\dag }]=1$, $[a_{\xi },a_{\eta }^{\dag }]=[a_{\eta
},a_{\xi }^{\dag }]=[a_{\eta },a_{\xi }]=[a_{\eta }^{\dag },a_{\xi }^{\dag
}]=0$. The Hamiltonian is%
\begin{equation*}
H=\Omega \left( a_{\xi }^{\dag }a_{\xi }+\frac{1}{2}\right) -\tilde{\Omega}%
\left( a_{\eta }^{\dag }a_{\eta }+\frac{1}{2}\right) .
\end{equation*}%
We see that $a_{\eta }^{\dag }$ are the creation operators of the fakeons
minus. The physical subspace $V$ is obtained by projecting them away.
Specifically, the vacuum $|0\rangle $ satisfies $a_{\xi }|0\rangle =a_{\eta
}|0\rangle =0$. The space $V$ is made of the states $|v\rangle $ such that $%
a_{\eta }|v\rangle =0$. It is generated by $(a_{\xi }^{\dag })^{n}|0\rangle $
and contains the wave functions of the form 
\begin{equation*}
\psi (q,Q)=\psi (\xi )\exp \left( -\frac{m\tilde{\Omega}\eta ^{2}}{2}\right)
.
\end{equation*}%
The reduced Hamiltonian reads $H_{V}=\Omega a_{\xi }^{\dag }a_{\xi
}+(1/2)(\Omega -\tilde{\Omega})$ and is bounded from below in $V$. Note that 
$H_{V}$ must be shifted by a constant to have a regular limit $\tau
\rightarrow 0$.

\section{The classicization of quantum gravity}

\setcounter{equation}{0}\label{class}

In this section we study the fakeon projection in the classical limit of
quantum gravity. For simplicity, we work at $\Lambda _{C}=0$, the
generalization to $\Lambda _{C}\neq 0$ being straightforward. The field
equations derived from the interim classical action (\ref{SQG2}) read%
\begin{eqnarray}
&&R^{\mu \nu }-\frac{1}{2}g^{\mu \nu }R=\frac{\kappa ^{2}}{\zeta }\left[ 
\mathrm{e}^{3\kappa \phi }fT_{\mathfrak{m}}^{\mu \nu }(\tilde{g}\mathrm{e}%
^{\kappa \phi },\Phi )+fT_{\phi }^{\mu \nu }(\tilde{g},\phi )+T_{\chi }^{\mu
\nu }(g,\chi )\right] ,  \notag \\
-\frac{1}{\sqrt{-\tilde{g}}} &&\partial _{\mu }\left( \sqrt{-\tilde{g}}%
\tilde{g}^{\mu \nu }\partial _{\nu }\phi \right) -\frac{m_{\phi }^{2}}{%
\kappa }\left( \mathrm{e}^{\kappa \phi }-1\right) \mathrm{e}^{\kappa \phi }=%
\frac{\kappa \mathrm{e}^{3\kappa \phi }}{3\zeta }T_{\mathfrak{m}}^{\mu \nu }(%
\tilde{g}\mathrm{e}^{\kappa \phi },\Phi )\tilde{g}_{\mu \nu },  \label{3eom}
\\
&&\frac{1}{\sqrt{-g}}\frac{\delta S_{\chi }(g,\chi )}{\delta \chi _{\mu \nu }%
}=\mathrm{e}^{3\kappa \phi }fT_{\mathfrak{m}}^{\mu \nu }(\tilde{g}\mathrm{e}%
^{\kappa \phi },\Phi )+fT_{\phi }^{\mu \nu }(\tilde{g},\phi ),  \notag
\end{eqnarray}%
where $T_{A}^{\mu \nu }(g)=-(2/\sqrt{-g})(\delta S_{A}(g)/\delta g_{\mu \nu
})$ are the energy-momentum tensors ($A=\mathfrak{m}$, $\phi $, $\chi $) and 
$f=\sqrt{\det \tilde{g}_{\rho \sigma }/\det g_{\alpha \beta }}$.

The projection onto the right subspace of configurations works as follows.
Solve the third equations of (\ref{3eom}) for $\chi _{\mu \nu }$ by means of
the half sum of the retarded and advanced Green functions. Then insert the
solution $\langle \chi _{\mu \nu }\rangle $ into the other two equations. So
doing, $\chi $ becomes a classical fakeon and the first two lines of (\ref%
{3eom}) with $\chi _{\mu \nu }\rightarrow \langle \chi _{\mu \nu }\rangle $
become the projected equations of the GSF\ theory. They are also the field
equations of the finalized classical action 
\begin{equation}
\mathcal{S}_{\text{QG}}^{\text{GSF}}(g,\phi ,\Phi )=S_{\text{H}}(g)+S_{\chi
}(g,\langle \chi \rangle )+S_{\phi }(\bar{g},\phi )+S_{\mathfrak{m}}(\bar{g}%
\mathrm{e}^{\kappa \phi },\Phi ),  \label{sgsf}
\end{equation}%
where~$\bar{g}_{\mu \nu }=g_{\mu \nu }+2\langle \chi _{\mu \nu }\rangle $.

If we want to treat $\phi $ as a fakeon as well (GFF\ theory), we solve the
second and third equations of (\ref{3eom}) for $\phi $ and $\chi _{\mu \nu }$
by means of the half sums of the retarded and advanced Green functions and
insert the solutions $\langle \phi \rangle $, $\langle \chi _{\mu \nu
}\rangle $ into the first equation. The finalized classical action is then 
\begin{equation}
\mathcal{S}_{\text{QG}}^{\text{GFF}}(g,\Phi )=S_{\text{H}}(g)+S_{\chi
}(g,\langle \chi \rangle )+S_{\phi }(\bar{g},\langle \phi \rangle )+S_{%
\mathfrak{m}}(\bar{g}\mathrm{e}^{\kappa \langle \phi \rangle },\Phi ).
\label{sgff}
\end{equation}

We can make these operations more explicit by expanding around flat space.
The field equations of the action (\ref{SQG}) can be written in the form%
\begin{equation}
\left( \zeta +\alpha \nabla ^{2}\right) G_{\mu \nu }+\frac{\alpha -\xi }{3}%
\left( \nabla _{\mu }\nabla _{\nu }-g_{\mu \nu }\nabla ^{2}\right) G=\kappa
^{2}T_{\mu \nu },  \label{eom}
\end{equation}%
where $G_{\mu \nu }$ is the Einstein tensor and $G=g^{\mu \nu }G_{\mu \nu }$
denotes its trace, while 
\begin{equation}
\kappa ^{2}T_{\mu \nu }\equiv \kappa ^{2}T_{\mathfrak{m}\mu \nu }+\frac{%
\alpha }{2}g_{\mu \nu }R^{\rho \sigma }R_{\rho \sigma }-2\alpha R_{\mu \rho
\nu \sigma }R^{\rho \sigma }+\frac{2\alpha +\xi }{3}R\left( R_{\mu \nu }-%
\frac{1}{4}g_{\mu \nu }R\right) .  \label{tmn}
\end{equation}

Now, write $g_{\mu \nu }=\eta _{\mu \nu }+2\kappa h_{\mu \nu }$, where $\eta
_{\mu \nu }$ is the flat-space metric, and decompose $G_{\mu \nu }$ as $%
G_{\mu \nu }^{0}+(\kappa ^{2}J_{\mu \nu }/\zeta )$, where $G_{\mu \nu }^{0}$
denotes its linear part in $h_{\mu \nu }$. We understand that the indices of 
$\eta _{\mu \nu }$, $\partial _{\mu }$, $h_{\mu \nu }$ and $G_{\mu \nu }^{0}$
are raised and lowered by means of $\eta _{\mu \nu }$. Split the left-hand
side of the field equation (\ref{eom}) into its linear part plus the rest.
Precisely, recalling that $\partial ^{\mu }G_{\mu \nu }^{0}=0$, write 
\begin{equation}
\left( \zeta +\alpha \nabla ^{2}\right) G_{\mu \nu }+\frac{\alpha -\xi }{3}%
\left( \nabla _{\mu }\nabla _{\nu }-g_{\mu \nu }\nabla ^{2}\right) G\equiv 
\mathbb{Q}_{\mu \nu }^{\phantom{\mu\nu}\rho \sigma }G_{\rho \sigma
}^{0}+\kappa ^{2}U_{\mu \nu },  \label{usi}
\end{equation}%
where $U_{\mu \nu }$ collects the corrections that are at least quadratic in 
$h_{\mu \nu }$ and%
\begin{equation}
\mathbb{Q}_{\mu \nu }^{\phantom{\mu\nu}\rho \sigma }\equiv (\zeta +\alpha
\partial ^{2})\mathbb{I}_{\mu \nu }^{\phantom{\mu\nu}\rho \sigma }-\frac{%
\alpha -\xi }{3}\partial ^{2}\pi _{\mu \nu }\pi ^{\rho \sigma }.  \label{q}
\end{equation}%
Here, $\partial ^{2}=\partial ^{\mu }\partial _{\mu }$ and $\mathbb{I}_{\mu
\nu }^{\phantom{\mu\nu}\rho \sigma }$ is the identity operator for
transverse symmetric tensors with two indices, while $\pi _{\mu \nu }$ is
the spin-1 projector:%
\begin{equation*}
\mathbb{I}_{\mu \nu }^{\phantom{\mu\nu}\rho \sigma }=\frac{1}{2}(\pi _{\mu
}^{\rho }\pi _{\nu }^{\sigma }+\pi _{\mu }^{\sigma }\pi _{\nu }^{\rho
}),\qquad \pi _{\mu \nu }=\eta _{\mu \nu }-\frac{\partial _{\mu }\partial
_{\nu }}{\partial ^{2}}.
\end{equation*}%
The operators $\mathbb{I}_{\mu \nu }^{\phantom{\mu\nu}\rho \sigma }$ and $%
\mathbb{Q}_{\mu \nu }^{\phantom{\mu\nu}\rho \sigma }$ satisfy obvious
symmetry properties and are transverse: $\partial ^{\mu }\mathbb{I}_{\mu \nu
}^{\phantom{\mu\nu}\rho \sigma }=\partial ^{\mu }\mathbb{Q}_{\mu \nu }^{%
\phantom{\mu\nu}\rho \sigma }=0$. Clearly, $\mathbb{I}_{\mu \nu }^{%
\phantom{\mu\nu}\alpha \beta }\mathbb{I}_{\alpha \beta }^{\phantom{\mu\nu}%
\rho \sigma }=\mathbb{I}_{\mu \nu }^{\phantom{\mu\nu}\rho \sigma }$. The
inverse of $\mathbb{Q}_{\mu \nu }^{\phantom{\mu\nu}\rho \sigma }$ is 
\begin{equation*}
\mathbb{P}_{\mu \nu }^{\phantom{\mu\nu}\rho \sigma }=\frac{1}{\zeta +\alpha
\partial ^{2}}\left( \mathbb{I}_{\mu \nu }^{\phantom{\mu\nu}\rho \sigma }+%
\frac{\alpha -\xi }{3}\frac{\partial ^{2}}{\zeta +\xi \partial ^{2}}\pi
_{\mu \nu }\pi ^{\rho \sigma }\right) ,
\end{equation*}%
i.e. $\mathbb{P}_{\mu \nu }^{\phantom{\mu\nu}\alpha \beta }\mathbb{Q}%
_{\alpha \beta }^{\phantom{\mu\nu}\rho \sigma }=\mathbb{Q}_{\mu \nu }^{%
\phantom{\mu\nu}\alpha \beta }\mathbb{P}_{\alpha \beta }^{\phantom{\mu\nu}%
\rho \sigma }=\mathbb{I}_{\mu \nu }^{\phantom{\mu\nu}\rho \sigma }$.

The projected equations of the GFF theory can be obtained by using (\ref{usi}%
) in (\ref{eom}) and inverting $\mathbb{Q}_{\mu \nu }^{\phantom{\mu\nu}\rho
\sigma }$. This gives%
\begin{equation}
G_{\mu \nu }^{0}=\kappa ^{2}\mathbb{P}_{\mu \nu }^{\phantom{\mu\nu}\rho
\sigma }(T_{\rho \sigma }-U_{\rho \sigma }).  \notag
\end{equation}%
Using the transversality of $\kappa ^{2}(T_{\mu \nu }-U_{\mu \nu })=\mathbb{Q%
}_{\mu \nu }^{\phantom{\mu\nu}\rho \sigma }G_{\rho \sigma }^{0}$ and
inserting the classical fakeon prescription (\ref{fprc}), we obtain%
\begin{equation}
G_{\mu \nu }^{0}=\frac{1}{\bar{M}_{\text{Pl}}^{2}}\left\langle T_{\mu \nu
}-U_{\mu \nu }+\frac{r_{\phi \chi }}{3}\left( \eta _{\mu \nu }\partial
^{2}-\partial _{\mu }\partial _{\nu }\right) \langle T-U\rangle _{\phi
}\right\rangle _{\chi },  \label{peq}
\end{equation}%
where $T=\eta ^{\mu \nu }T_{\mu \nu }$, $U=\eta ^{\mu \nu }U_{\mu \nu }$, $%
r_{\phi \chi }=(m_{\phi }^{2}-m_{\chi }^{2})/(m_{\phi }^{2}m_{\chi }^{2})$
and the average $\langle \cdots \rangle _{F}$ associated with the fakeon $F$
of mass $m_{F}$ is defined as%
\begin{equation}
\langle \mathcal{O}\rangle _{F}\equiv \frac{m_{F}^{2}}{2}\left[ \left. \frac{%
1}{m_{F}^{2}+\partial ^{2}}\right\vert _{\text{ret}}+\left. \frac{1}{%
m_{F}^{2}+\partial ^{2}}\right\vert _{\text{adv}}\right] \mathcal{O}.
\label{of}
\end{equation}

In covariant form, the field equations of the GFF theory read%
\begin{equation}
R_{\mu \nu }-\frac{1}{2}g_{\mu \nu }R=\frac{1}{\bar{M}_{\text{Pl}}^{2}}%
T_{\mu \nu }^{\text{GFF}},  \label{feom}
\end{equation}%
where%
\begin{equation*}
T_{\mu \nu }^{\text{GFF}}=\left\langle T_{\mu \nu }-U_{\mu \nu }+\frac{%
r_{\phi \chi }}{3}\left( \eta _{\mu \nu }\partial ^{2}-\partial _{\mu
}\partial _{\nu }\right) \langle T-U\rangle _{\phi }\right\rangle _{\chi
}+J_{\mu \nu }.
\end{equation*}%
Note that although $T_{\mu \nu }^{\text{GFF}}$ is not manifestly covariant,
it can be put in fully covariant form, because equations (\ref{feom}) are
equivalent to the field equations of (\ref{sgff}), which are covariant.
Since (\ref{feom}) is a perturbative version of the field equations of (\ref%
{sgff}), the covariantization can be achieved iteratively, starting by
replacing $\partial ^{2}$ with its covariant version in the definition (\ref%
{of}) of the fakeon average.

Now we consider the GSF\ theory. Tracing (\ref{peq}) with the flat-space
metric and using the definition (\ref{of}), we obtain%
\begin{equation*}
\frac{r_{\phi \chi }}{\bar{M}_{\text{Pl}}^{2}}\langle \langle T-U\rangle
_{\phi }\rangle _{\chi }=\frac{1}{\bar{M}_{\text{Pl}}^{2}m_{\chi }^{2}}%
\langle T-U\rangle _{\chi }-\frac{1}{m_{\phi }^{2}}G^{0},
\end{equation*}%
where $G^{0}=G_{\mu \nu }^{0}\eta ^{\mu \nu }$. Inserting this formula back
into (\ref{peq}), we also get%
\begin{equation*}
G_{\mu \nu }^{0}+\frac{1}{3m_{\phi }^{2}}\left( \eta _{\mu \nu }\partial
^{2}-\partial _{\mu }\partial _{\nu }\right) G^{0}=\frac{1}{\bar{M}_{\text{Pl%
}}^{2}}\left( \langle T_{\mu \nu }-U_{\mu \nu }\rangle _{\chi }+\frac{1}{%
3m_{\chi }^{2}}\left( \eta _{\mu \nu }\partial ^{2}-\partial _{\mu }\partial
_{\nu }\right) \langle T-U\rangle _{\chi }\right) .
\end{equation*}%
Then, calling 
\begin{equation*}
K_{\mu \nu }=\frac{\bar{M}_{\text{Pl}}^{2}}{3m_{\phi }^{2}}\left[ \left(
g_{\mu \nu }\nabla ^{2}-\nabla _{\mu }\nabla _{\nu }\right) G-\left( \eta
_{\mu \nu }\partial ^{2}-\partial _{\mu }\partial _{\nu }\right) G^{0}\right]
,
\end{equation*}%
we find the projected field equations of the GSF theory, which read%
\begin{equation}
G_{\mu \nu }+\frac{1}{3m_{\phi }^{2}}\left( g_{\mu \nu }\nabla ^{2}-\nabla
_{\mu }\nabla _{\nu }\right) G=\frac{1}{\bar{M}_{\text{Pl}}^{2}}T_{\mu \nu
}^{\text{GSF}},  \label{gfeq}
\end{equation}%
where%
\begin{equation*}
T_{\mu \nu }^{\text{GSF}}=\langle T_{\mu \nu }-U_{\mu \nu }\rangle _{\chi }+%
\frac{1}{3m_{\chi }^{2}}\left( \eta _{\mu \nu }\partial ^{2}-\partial _{\mu
}\partial _{\nu }\right) \langle T-U\rangle _{\chi }+J_{\mu \nu }+K_{\mu \nu
}.
\end{equation*}%
Again, the tensor $T_{\mu \nu }^{\text{GSF}}$ can be iteratively
covariantized to put (\ref{gfeq}) in a fully covariant form.

The left-hand side of equation (\ref{gfeq}) coincides with the left-hand
side of (\ref{eom}), divided by $\zeta $, in the limit $\alpha \rightarrow 0$%
, i.e. $m_{\chi }\rightarrow \infty $. Moreover, the right-hand side of (\ref%
{gfeq}) contains only the average $\langle \cdots \rangle _{\chi }$. These
facts show that the field $\chi _{\mu \nu }$ is integrated out. It is easy
to prove that the surviving degrees of freedom are the graviton and the
massive scalar $\phi $, and that their poles have positive residues.

Instead, the left-hand side of equation (\ref{feom}) coincides with the
left-hand side of (\ref{eom}), divided by $\zeta $, in the limit where both $%
m_{\chi }$ and $m_{\phi }$ are sent to infinity. The right-hand side of (\ref%
{feom}) contains both the averages $\langle \cdots \rangle _{\chi }$ and $%
\langle \cdots \rangle _{\phi }$, which means that both $\chi _{\mu \nu }$
and $\phi $ are integrated out.

The projected equations (\ref{feom}) and (\ref{gfeq}) are the classical
field equations of quantum gravity. They are written so that the initial
conditions, or the boundary conditions, are those dictated by their
left-hand sides. They are perturbative in $\kappa $ (but not in $\alpha $
and $\xi $). The cosmological constant can be reinstated by replacing $%
T_{\mu \nu }$ with $T_{\mu \nu }+g_{\mu \nu }(\Lambda _{C}/\kappa ^{2})$.

The averages $\langle \cdots \rangle _{\chi }$ and $\langle \cdots \rangle
_{\phi }$ that appear on the right-hand sides of the equations are the main
remnants of the classicization. They show that the violations of
microcausality and their intrinsic nonlocalities survive the classical limit
of quantum gravity.

By searching for exact solutions of physical interest and comparing them
with the solutions of the Einstein equations, we may identify complex
systems and nonperturbative configurations where the effects of the
violations of microcausality get amplified enough to become detectable. The
masses $m_{\chi }$ and $m_{\phi }$ might be sufficiently small to let us
uncover the first signs of quantum gravity without having to reach the
Planck scale.

\section{Conclusions}

\setcounter{equation}{0}\label{conclu}

In this paper we have studied the properties of the fakeons and their role
in quantum gravity. When fakeons are present, the quantization process
includes an additional step, since the starting local action is just an
interim one. The finalized classical action emerges from the classicization
of the quantum theory.

A fakeon $\chi _{\mu \nu }$ of spin 2 and a scalar field $\phi $ are able to
make quantum gravity renormalizable while preserving unitarity. Depending on
whether $\phi $ is physical or fake, we have two possibilities, the GSF and
GFF theories. We worked out the finalized classical actions in the two
cases, which are (\ref{sgsf}) and (\ref{sgff}), respectively. The classical
field equations derived from them are (\ref{gfeq}) and (\ref{feom}). Among
the other things, the results make clear that the violations of
microcausality survive the classical limit.

The corrections to general relativity become important at energies
comparable with the masses $m_{\chi }$ and $m_{\phi }$ of $\chi _{\mu \nu }$
and $\phi $. The values of such masses in nature could be much smaller than
the Planck mass and still be compatible with every experimental observations
made so far. Yet, those values might still be too large to detect new
effects in scattering processes and other elementary or perturbative
phenomena. For this reason, it may be interesting to study the classicized
actions derived here, because they give us the chance to investigate
collective, nonperturbative effects. A possibility is to study how the
solutions of the field equations of general relativity get modified and
search for situations where the violations of microcausality get amplified
enough to become detectable. This kind of investigation might also help us
put experimental bounds on the values of $m_{\chi }$ and $m_{\phi }$.

It is not easy to detect the violations of causality directly, since when we
solve a self-consistent system (no external sources being involved), we know
in advance what the interactions will be in the future (as functions of the
fields), which makes it hard to discriminate what is expected from what is
unexpected. However, the corrections to general relativity predicted by the
equations (\ref{feom}) and (\ref{gfeq}) might help us detect the violations
indirectly or test other nontrivial predictions of quantum gravity, maybe
from the observations of black holes or by studying the consequences on
cosmology.

\section*{Acknowledgments}

We are grateful to U. Aglietti, L. Bracci, M. Piva and A. Strumia for useful
discussions.

\end{document}